 \documentclass[final,5p,times,twocolumn]{elsarticle}

\usepackage{graphicx}
\usepackage{dcolumn}
\usepackage{bm}
\usepackage[colorlinks=true, allcolors=blue]{hyperref}
\usepackage{xcolor}
\usepackage{ulem}
\usepackage{comment}
\usepackage{refcount}

\newcommand{\be}{\begin{equation}}
\newcommand{\ee}{\end{equation}}
\newcommand{\ba}{\begin{eqnarray}}
\newcommand{\ea}{\end{eqnarray}}

\journal{Physics Letters B}

\begin{document}

\begin{frontmatter}

\title{Projective Imaging of High-Energy Nuclei via Coherent Exclusive Vector Meson Production in Electron-Nucleus Collisions}


\author[first]{Maci Kesler}
\author[first]{Ashik Ikbal Sheikh}
\author[second]{Rongrong Ma}
\author[second]{Zhoudunming Tu}
\author[second]{Thomas Ullrich}
\author[first,second]{Zhangbu Xu}

\affiliation[first]{organization={Department of Physics, Kent State University},
            city={Kent},
            postcode={44242}, 
            state={OH},
            country={USA}}

\affiliation[second]{organization={Physics Department, Brookhaven National Laboratory},
            city={Upton},
            postcode={11973}, 
            state={NY},
            country={USA}}


\begin{abstract}

One of the major goals of modern nuclear experiments is to study the distributions of gluons inside nuclei at high energy. A key measurement is the coherent exclusive vector meson (VM) production in diffractive electron-nucleus collisions, where the gluon spatial distribution inside the nucleus can be obtained through a Fourier transform of the squared nuclear momentum transfer ($|t|$) distribution.  This research aims to overcome the two main obstacles of the $|t|$ measurement: limited precision in measuring $|t|$ arising from the momentum resolution of the outgoing electron and the overwhelming incoherent background. We demonstrate that by measuring the projected $|t|$ distribution along the direction perpendicular to the electron scattering plane, the effect of the outgoing electron's momentum resolution can be effectively mitigated, and the diffractive pattern is largely restored. Furthermore, we propose to measure the angular distribution of the VM's decay daughters to statistically remove the incoherent background. 
 
\end{abstract}


\end{frontmatter}


\section{Introduction}
\label{sec:intro}

\begin{figure}
    \includegraphics[scale=0.85]{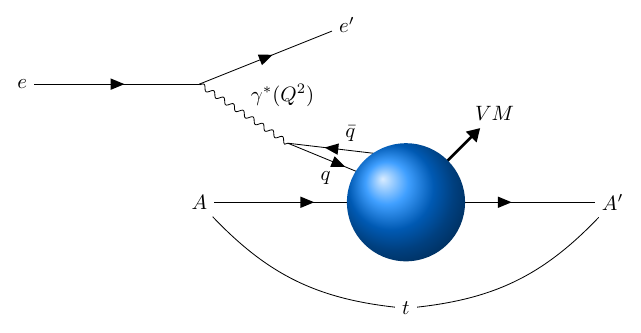}
    \caption
    {
     Exclusive VM production in a coherent diffractive $e$+$A$ collision.
    }
    \label{fig:diagram_VM_production}
\end{figure}

A central objective of nuclear physics experiments is to image the spatial distribution of gluons within nuclei and to understand underlying dynamics, such as gluon saturation~\cite{Accardi:2012qut,osti_1765663,AbdulKhalek:2021gbh,Anderle:2021wcy,agostini2021large,LHeC2022} at various proposed facilities in US, China, and Europe. A critical measurement is the coherent production of exclusive vector mesons (VMs), such as $\rho^{0}$, $\omega$, $\phi$, and J$/\psi$, etc, in diffractive electron-nucleus ($e$+$A$) collisions \cite{Klein_1999,Caldwell:2010zza,Toll:2012mb,Lomnitz:2018juf,Krelina:2019gee} as depicted in Fig. \ref{fig:diagram_VM_production}. An incoming electron emits a virtual photon ($\gamma^{*}$) with virtuality $Q^{2}$, which then scatters off the nucleus. Specifically, the virtual photon fluctuates into a quark-antiquark pair, interacts with the nucleus through an exchange of a Pomeron, and emerges as a VM without any other final-state particles produced in the process. The incoming nucleus remains intact after scattering. The squared four-momentum transfer of the nucleus is denoted as $t$, and the spatial distribution of gluons inside the nucleus can be extracted through a Fourier transformation of the $|t|$ distribution in such coherent events. For cases where the nucleus breaks up, they are referred to as incoherent interactions.

Experiments~\cite{Lanzerotti:1968PhysRev.166.1365} in 1968 showed for the first time diffraction-like coherent photoproduction of $\rho^0$ using 5.5 GeV and 3 GeV photon beams on proton and carbon targets. Subsequently, exclusive coherent photonuclear production of $\rho^0$ was used at various facilities to determine the nuclear radii in the early 1970s ~\cite{Alvensleben:1970uw,Alvensleben:1970uv,TingPhysRevLett.27.888,CornellPhysRevD.4.2683} with photons of energies $\sim\mathcal{O}$(GeV)  incident on nuclear targets. The $|t|$ distribution from the diffractive process was fitted with the Fourier-Transform of the Woods-Saxon nuclear density distribution~\cite{Alvensleben:1970uv,Mcclellan:1971rxw}. Although these experiments showed success of this technique, the approach is limited to low luminosity and low incident photon energy~\cite{Brandenburg:2025one}. The cross section ratios between coherent and incoherent interactions for photonuclear $\rho^0$ production have also been used as an effective tool to study the color transparency at HERA and JLab in later experiments~\cite{JAINColor199667,CLASELFASSI2012326}. While similar coherent photonuclear production of VMs in ultraperipheral heavy-ion collisions \cite{ALICE:2020ugp,STAR:2017enh,STAR:2022wfe,STAR:2023nos,Brandenburg:2025one,Duan:2024,Hagiwara:2021xkf} can also be used to study gluon distribution in a heavy nucleus, it is limited to the phase space of $Q^2 \simeq 0$ while integrating over a range of photon kinematics. Measuring the VM production in an $e$+$A$ collider allows access to a large $x-Q^2$ range with precise kinematics, gaining a more comprehensive understanding of the gluon distributions in nuclei. Here, $x$ denotes the fraction of the nucleon momentum carried by the struck parton.

Several experiments at different facilities, such as JLab, HERA, etc. \cite{2008arXiv0812.0539B,H1:2009cml,paolone_2014,paolone_2016}, measure coherent VM production at low $|t|$ with limited success. The difficulty arises from two critical challenges:
(i) limited precision in measuring $|t|$ arising mainly from the momentum resolution of the outgoing electron, and (ii) the overwhelming incoherent background. These measurements highlighted as part of a high-priority physics program by the National Academies of Sciences, Engineering, and Medicine in their report ``An Assessment of U.S.-Based Electron-Ion Collider Science" (Fig. 2.9~\cite{NAP25171}) and were prominently featured in the Electron-Ion Collider (EIC) white paper~\cite{Accardi:2012qut}. However, detector simulations, initially by the EIC proto-collaborations ECCE and ATHENA, and later by ePIC~\cite{Adkins:2022jfp,ATHENA:2022hxb}, revealed that even with the highest achievable resolution in current detector designs, the diffractive pattern for coherent VM production remains unresolved. The finite detector resolution smears the signal to such an extent that the diffractive structure is effectively washed out and becomes unobservable~\cite{ATHENA:2022hxb}.

In this work, we exploit the decomposition of $|t|$ into three components and select a specific phase space to enhance the diffractive pattern without losing critical information. The $|t|$ distribution is projected along the normal direction of the electron scattering plane to alleviate the effect of electron momentum resolution and to a large extent restore the expected features of the $|t|$ distribution for coherent events. We also propose to utilize transversely polarized electron beams and take advantage of the unique angular distribution of the daughter particles from VM decays to determine the contribution of incoherent production on an ensemble basis. We will then use the kinematics and detector design of the EIC \cite{Accardi:2012qut,AbdulKhalek:2021gbh}, to be built at the Brookhaven National Laboratory (BNL) in thue. 
\section{Method}
\label{sec:method}

For a heavy nucleus, the $|t|$ distribution is given by a form factor, whose analytic form can be obtained through the convolution of the Yukawa potential with the Woods-Saxon distribution approximated as a hard sphere potential \cite{Lomnitz:2018juf,Toll:2012mb}:
\be
F(t) = \frac{4 \pi \rho_0}{A q^3}\left[\sin\left(t R\right) - t R \cos\left(t R\right)\right] \left(\frac{1}{1+a^2t^2}\right),
\label{eq:true_form_factor}
\ee
where $t = q^2$, $\rho_0$ is the nuclear density at the center of the nucleus, $A$ is the atomic mass number, $R$ is the nuclear radius, and the range of the Yukawa potential is given by $a = 0.7$ fm.

\subsection{Current Challenges}
\label{subsec:current.challenges}
\begin{figure}
    \includegraphics[scale=0.44]{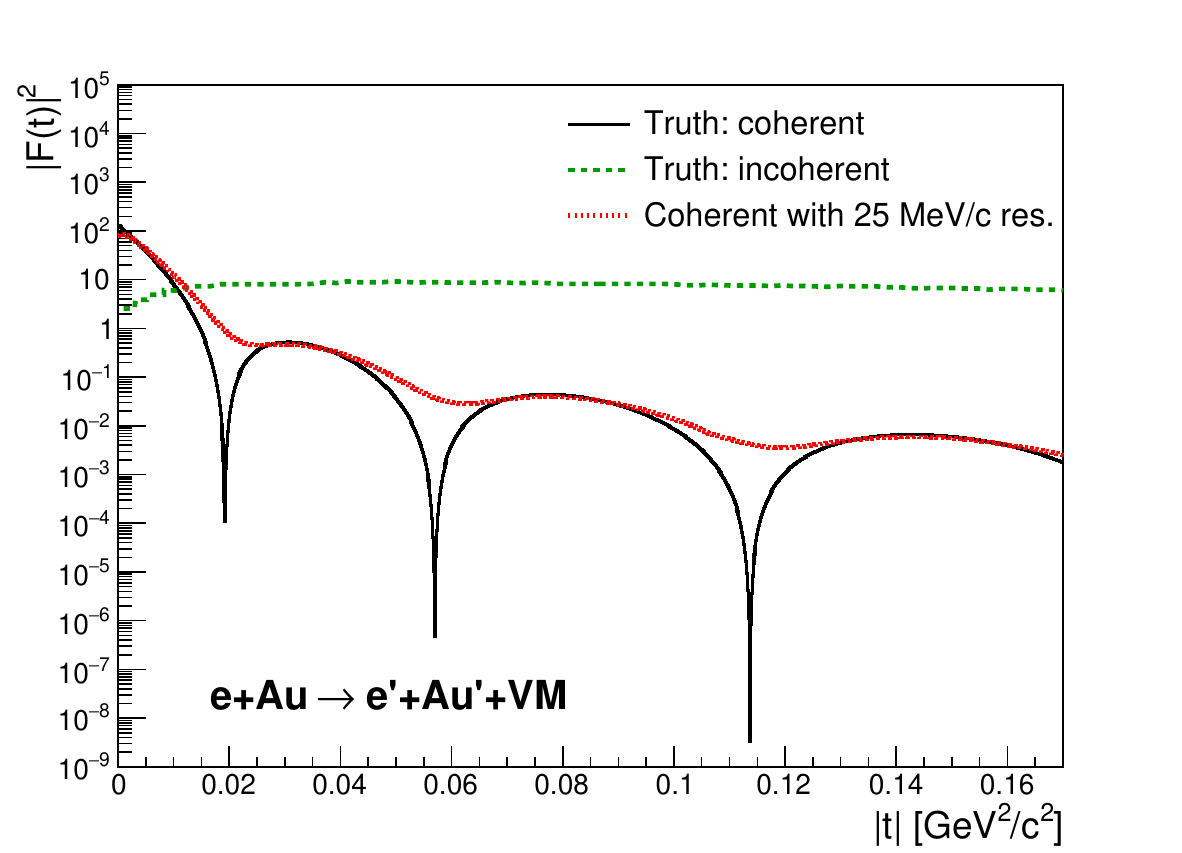}
    \caption
    {$|t|$ distributions for coherent and incoherent exclusive VM production in $e$+Au collisions. The coherent distribution, based on Eq. \ref{eq:true_form_factor}, is shown as the black solid curve. This is the maximum possible diffraction pattern one can achieve. The red dotted line represents the expected $|t|$ distribution with a resolution of 25 MeV$/c$ \cite{osti_1765663}. The incoherent production for J/$\psi$ \cite{ATHENA:2022hxb} (green dashed line) is also displayed as an example. 
    }
    \label{fig:coherent_incoherent_diffraction}
\end{figure}

The challenges in measuring the $|t|$ distribution are illustrated in Fig.~\ref{fig:coherent_incoherent_diffraction}. The black solid curve shows the typical $|t|$ distribution for coherently produced VMs in $e$+Au collisions based on Eq. \ref{eq:true_form_factor}. A distinct structure of peaks and valleys is seen, where the positions of the valleys or minima are determined by the gluon spatial distribution in the scattered nucleus, while the magnitude of the distribution is sensitive to saturation effects. The expected $|t|$ distribution after considering a resolution of 25 MeV/$c$ as inspired by the EIC detector design \cite{osti_1765663}, mainly driven by the outgoing electron’s momentum resolution, is shown as a red dotted curve. The structure of peaks and valleys is largely washed out, making it extremely difficult to extract the spatial gluon distribution. In addition to coherently produced VMs, there are also VMs produced in incoherent diffractive events. The green dashed line in Fig.~\ref{fig:coherent_incoherent_diffraction} \cite{ATHENA:2022hxb} shows  as an example incoherent J/$\psi$ meson production, whose yield dominates over coherent production in most of the interested $|t|$ range. The incoherent production is of interest on its own, but it acts as an overwhelming background for measuring coherent production. While the incoherent process can be suppressed experimentally by tagging nuclear fragments on an event-by-event basis  \cite{ATHENA:2022hxb}, making it possible to resolve the first minimum, resolving the second and third minima is extremely challenging, if not impossible.
%

\subsection{Projective Technique}
\label{subsec:projection.technique}

\subsubsection{Challenge I: limited resolution}

The squared four-momentum transfer, $|t|$, is defined as: 
\be
|t| = -(p_{A'} - p_A)^2,
\ee
where $p_A$ and $p_{A'}$ are the four momenta of the colliding nucleus before and after scattering. Since measuring $p_{A'}$ is experimentally challenging, invoking energy and momentum conservation yields:
\be
|t| = -(p_e - p_{e'} - p_{\rm VM})^2,
\ee
where $p_{\rm VM}$, $p_{e'}$, and $p_e$ are the four momenta for the VM, scattered electron and incoming electron. 
It is evident that the precision of $|t|$  determination is limited by 
the resolution in measuring scattered electron's momentum.  

\begin{figure*}
    \centering
    \includegraphics[width=\textwidth]{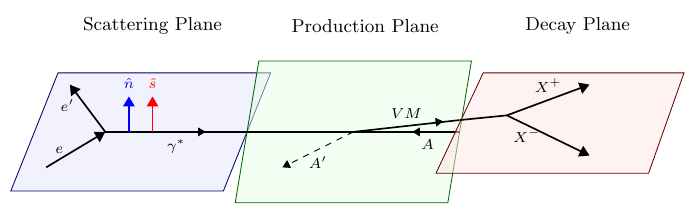}
    \caption
    {
    Coherent exclusive VM production in an $e$+$A$ collision. The normal direction ($\hat{n}$) to the electron scattering plane is shown as the blue vertical arrow. The spin direction ($\hat{s}$) of the virtual photon emitted by the incoming electron can align with the normal direction in specific configurations. See text for details.
    }
    \label{fig:t_reconstruction}
\end{figure*}

To overcome this challenge of insufficient experimental precision, we propose to measure the projection of $|t|$ along the normal direction of the electron scattering plane. Figure \ref{fig:t_reconstruction} shows another diagram for a coherent exclusive VM production event ($e$ + $A \rightarrow e^\prime$ + $A'$ + $VM$), where $e$ and $e^\prime$ denote incoming and outgoing electrons, $\gamma^*$ represents the emitted virtual photon, $A$ and $A'$ are the incoming and outgoing nuclei, and VM indicates the produced VM that decays into $X^+$ and $X^-$~\cite{Lomnitz:2018juf}. The blue vertical arrow points along the normal direction ($\hat{n}$) of the electron scattering plane, spanned by the momenta of incoming and outgoing electrons. Here, we define the $z$ direction as the virtual photon's travel direction within the electron scattering plane, the $y$ direction to be parallel to $\hat{n}$, and the $x$ direction within the electron scattering plane and perpendicular to $z$.


The projection of $|t|$ to the $\hat{n}$ direction is given by: 
\ba
|t_{\hat{n}}| &=& [(p_e - p_{e'} - p_{\rm VM}) \cdot \hat{n}]^2,
\label{eq:conservation_4momenta}
\ea
where $\hat{n}=(0,1,0,0)$. Taking into account that $\hat{n}$ is perpendicular to the momentum vectors of incoming and outgoing electrons, Eq. \ref{eq:conservation_4momenta} reduces to:
\ba
|t_{\hat{n}}| &=& (p_e \cdot \hat{n} - p_{e'} \cdot \hat{n} - p_{\rm VM} \cdot \hat{n})^2 \nonumber \\
&=& (p_{\rm  VM} \cdot \hat{n})^2.
\label{eq:t_project}
\ea
This effectively eliminates dependence on the electrons’ momenta, so their measurement resolution does not contribute
to $|t_{\hat{n}}|$. Only the electrons’ flying directions before and after scattering are needed, which can be measured with much better precision than their momenta. 


If one directly plots the form factor as a function of $|t_{\hat{n}}|$, equivalent to integrating over other components of $|t|$, smearing in the diffractive pattern would still appear. Instead, we have to find the phase space where $|t_{\hat{n}}|$ is the dominant component of $|t|$. To achieve that, we decompose $t$ as

\be
t = t_{\perp} + t_{\parallel},
\label{eq:t}
\ee
where $t_{\perp} = t_x +t_y$ and $t_{\parallel}$ are transverse and longitudinal components. $t_{\parallel}$ lies along the $z$ direction, while $t_x$ and $t_y$ are the components along $x$ and $y$ directions. $t_{\parallel}$ is determined by the VM mass and longitudinal momentum conservation, and is neglected in this work as usually $t_{\parallel}\ll t_{\perp}$ for high-energy nuclei \cite{Lomnitz:2018juf}.  Note that $t_y \equiv t_{\hat{n}}$, and $t_x$ is directly affected by the outgoing electron's momentum resolution. We further define:
\be
t_{\perp} = q_{\perp}^2 = q_x^2 + q_y^2,
\ee
and
\be
q_x = q_{\perp} \sin(\theta),\,\,\, q_y = q_{\perp} \cos(\theta),
\label{eq:parameterize_q}
\ee
where $\theta$ is the angle between $q_{\perp}$ and $q_y$ in the $x-y$ plane. 

\begin{figure}
    \includegraphics[scale=0.28]{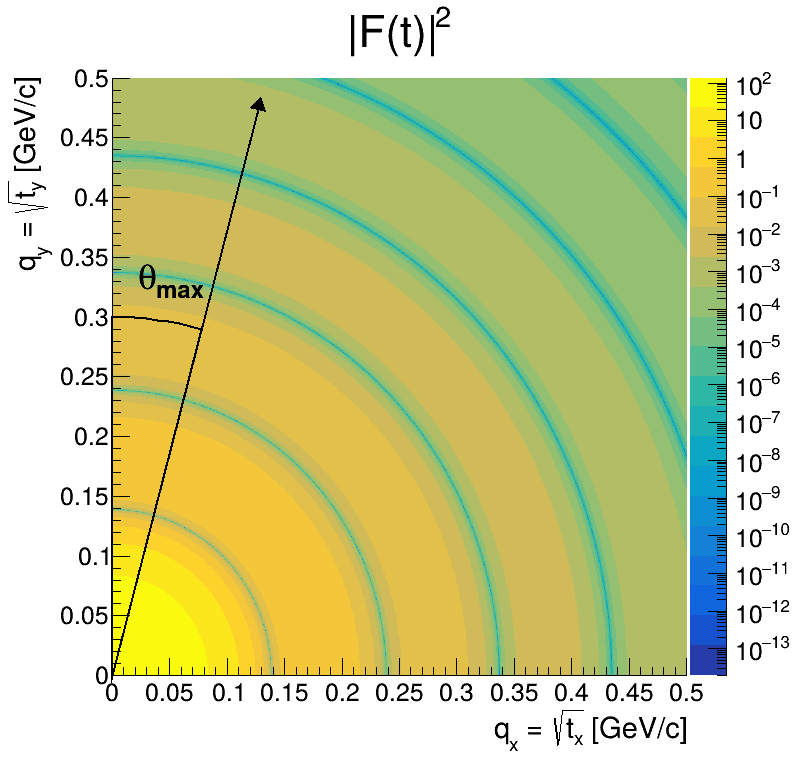}
    \caption
    {
    Distribution of the form factor from Eq. \ref{eq:true_form_factor} as a function of ($q_x$, $q_y$).
    }
    \label{fig:ffq_true}
\end{figure}

The form factor for coherent VM production (Eq. \ref{eq:true_form_factor}) as a function $q_x$ and $q_y$ is shown in Fig. \ref{fig:ffq_true}, where the diffractive pattern of peaks and valleys show up as concentric circles.
We can now apply an angle cut ($\theta < \theta_{\rm max}$) to select the phase space where $t_y$ dominates over $t_x$, approximating the projected $|t|$ distribution along the $\hat{n}$ or $y$ direction:
\be
|F(t)|^2 \rightarrow \int^{\theta_{\rm max}}_0 |F'(t,\theta)|^2 \, d\theta
= |F_{\hat{n}} (t)|^2.
\label{eq:t-projection}
\ee
Consequently, when $\theta_{\rm max}$ is small, most of the  $q_x$ component, and thus the outgoing electron's momentum resolution, is eliminated.

The effectiveness of the proposed approach in suppressing the impact of limited precision in measuring $t_x$ can be demonstrated analytically by expanding the form factor (Eq. \ref{eq:true_form_factor}) around $q_x = 0$ with a Taylor series. Specifically, 
\be
F(t)\approx F(t)|_{q_x=0} + \frac{\partial F}{\partial q_x}|_{q_x=0}\Delta q_x + \frac{\partial^2 F}{\partial q^2_x}|_{q_x=0}\frac{1}{2}(\Delta q_x)^2
\label{eq:1exp}
\ee
The first-order partial derivative is: 
\ba
\frac{\partial F}{\partial q_x}|_{q_x=0}&=&\left(\frac{\partial F}{\partial t}\frac{\partial t} {\partial q_x}\right)|_{q_x=0} \nonumber \\
&=&\left(\frac{\partial F}{\partial t}\times2q_x\right)|_{q_x=0} \nonumber \\
&=&0
\label{eq:2exp}
\ea
The second-order partial derivative is:
\ba
\frac{\partial^2 F}{\partial q_x^2}|_{q_x=0} &=& \frac{\partial}{\partial q_x} \frac{\partial F}{\partial q_x}|_{q_x=0} \nonumber \\
&=& \frac{\partial}{\partial q_x} \left(\frac{\partial F}{\partial t} \frac{\partial t}{\partial q_x}\right)|_{q_x=0}\nonumber\\
&=& \left[\frac{\partial}{\partial q_x} \left(\frac{\partial F}{\partial t}\right)\times 2 q_x + \frac{\partial F}{\partial t} \frac{\partial}{\partial q_x} (2 q_x)\right]|_{q_x=0} \nonumber\\
&=& 2 \frac{\partial F}{\partial t}|_{q_x=0}
\label{eq:3exp}
\ea
Thus:
\be
F(t) \approx F(q_y^2) + \frac{\partial F}{\partial t}|_{q_x=0}(\Delta q_x)^2
\ee
%
By projecting $|F(t)|$ to the $\hat{n}$ or $y$ direction, the resolution on $t_x$ appears at second order. We note that it is not uncommon that specific kinematics are defined by the convenience for certain physics objective~\cite{collins1997lightconevariablesrapidity,collins2011foundations}.

\subsubsection{Challenge II: incoherent background}

To overcome the challenge of the coherent signal being swamped by the incoherent background, we propose to exploit the decay pattern of the VM with respect to $\hat{n}$ to accurately determine the fraction of coherently produced VMs in collisions of transversely polarized electron beams, {\it i.e.}, the electron spin is perpendicular to its momentum.

In coherent events where the incoming electron flips its spin after scattering, the spin of the emitted virtual photon ($\hat{s}$ in Fig. \ref{fig:t_reconstruction})  will align with the spin of the incoming electron since the exchanged Pomeron has 0 spin, and finally transfers to the produced VM. In other words, the spin of the VM will align with $\hat{n}$ in such events. We can then project the momentum of the VM’s decay daughter onto the VM’s spin direction, $\hat{s}$ ({\it i.e.}, $\hat{n}$), a $\cos(2\phi)$ modulation is expected \cite{STAR:2022wfe}, where $\phi$ denotes the angle between the momentum of the decay daughter in the VM's rest frame and $\hat{n}$. On the other hand, if the spin of the incoming electron does not flip during scattering, there will be no preferred direction for the spin of the virtual photon or the VM, and a flat $\phi$ distribution is expected. Similarly, in an incoherent event where the nucleus breaks up, the spin of the VM is expected to be random with respect to $\hat{n}$, and therefore no $\cos(2\phi)$ modulation should be observed either. Consequently, the fraction of coherent events where the incoming electron flips its spin is equal to $\langle\cos(2\phi)\rangle$. Here, the average runs over all exclusive VM events. If we assume that the probability for the incoming electron to flip its spin is $C$, which is expected to be independent of $|t|$ since they are causally disconnected, the fraction of total coherent events will be $\langle\cos(2\phi)\rangle/C$ in the event sample. By measuring such a fraction in each $|t|$ bin, one can obtain the $|t|$ distribution for the coherent VM production on an ensemble basis. As a by-product, one can also obtain the incoherent VM cross section as a function of $|t|$, from which fluctuations in the gluon density can be studied.

To summarize, this projective imaging technique takes advantage of several features of diffractive processes:
\begin{itemize}
\item 
The interaction of the virtual photon and the Pomeron as shown in Fig.~\ref{fig:diagram_VM_production} is isotropic in the $t_{\perp}$ plane because the exchanged Pomeron is spin 0. This means that selection on a specific wedge of the $t_{\perp}$ plane does not lose any information. Measurements of higher-order interactions may be affected by the phase space selection due to non-spherical nuclear geometry and should be taken into account in future work~\cite{Duan:2024,Hagiwara:2021xkf,Blaizot:2025scr}. 
\item 
The $|t|$ resolution is dominated by the energy and momentum measurements of the scattered electrons, while their angular resolution is negligible. Therefore, a phase space selection perpendicular to the electron scattering plane minimizes the impact of the detector resolution.
\item 
The fact that the Pomeron has 0 spin allows the transfer of the electron polarization to the emitted photon and finally to the coherently produced VM, with the spin direction fully determined. Due to the conservation of angular momentum, the decay pattern of the VM can therefore be utilized to distinguish coherent from incoherent production.  
\end{itemize}

\section{Results}
\label{sec:results}

In this section, we apply the proposed projective technique and demonstrate that one is indeed able to alleviate the limited $|t|$ resolution and largely restore the diffractive pattern for coherent VM production by measuring the projective $|t|$ distribution in a specified phase space. The statistical separation of coherent and incoherent production through decay patterns is less intricate and will be further explored in future work. 

To simulate detector resolution, we smear the coherent $|t|$ distribution from its analytic form, {\it i.e.}, black solid curve in Fig. \ref{fig:coherent_incoherent_diffraction}, with a $|t|$-independent resolution of 25 MeV/$c$ as inspired by the EIC detector proposal \cite{ATHENA:2022hxb}. This is done by convoluting Eq. \ref{eq:true_form_factor} with a Gaussian distribution to smear the $q_x$ component, and the result is shown as red dotted curve in Fig. \ref{fig:coherent_incoherent_diffraction}.

\begin{figure}
    \includegraphics[scale=0.28]{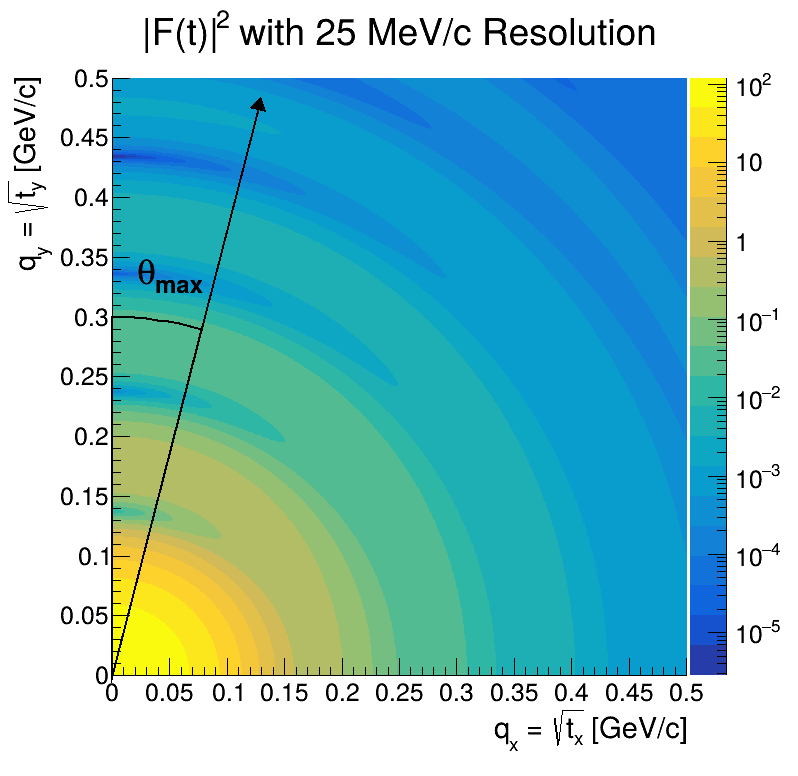}
    \caption
    {Distribution of the form factor from Eq. \ref{eq:true_form_factor} as a function of ($q_x$, $q_y$), with 25 MeV/$c$ resolution added to the $q_x$ axis. A wedge cut of $\theta_{\rm max} = \pi/12$ from the $q_y$ axis is indicated by arrow.
    }
    \label{fig:ffq_25MeV}
\end{figure}

The smeared $|F(t)|^{2}$ distribution as a function of $q_x$ and $q_y$ is shown in Fig. \ref{fig:ffq_25MeV}. Compared to Fig. \ref{fig:ffq_true} where no detector resolution is included, the distribution becomes more elliptical along the $q_x$ axis, a direct consequence of the electron momentum resolution. 
%

\begin{figure}
    \includegraphics[scale=0.47]{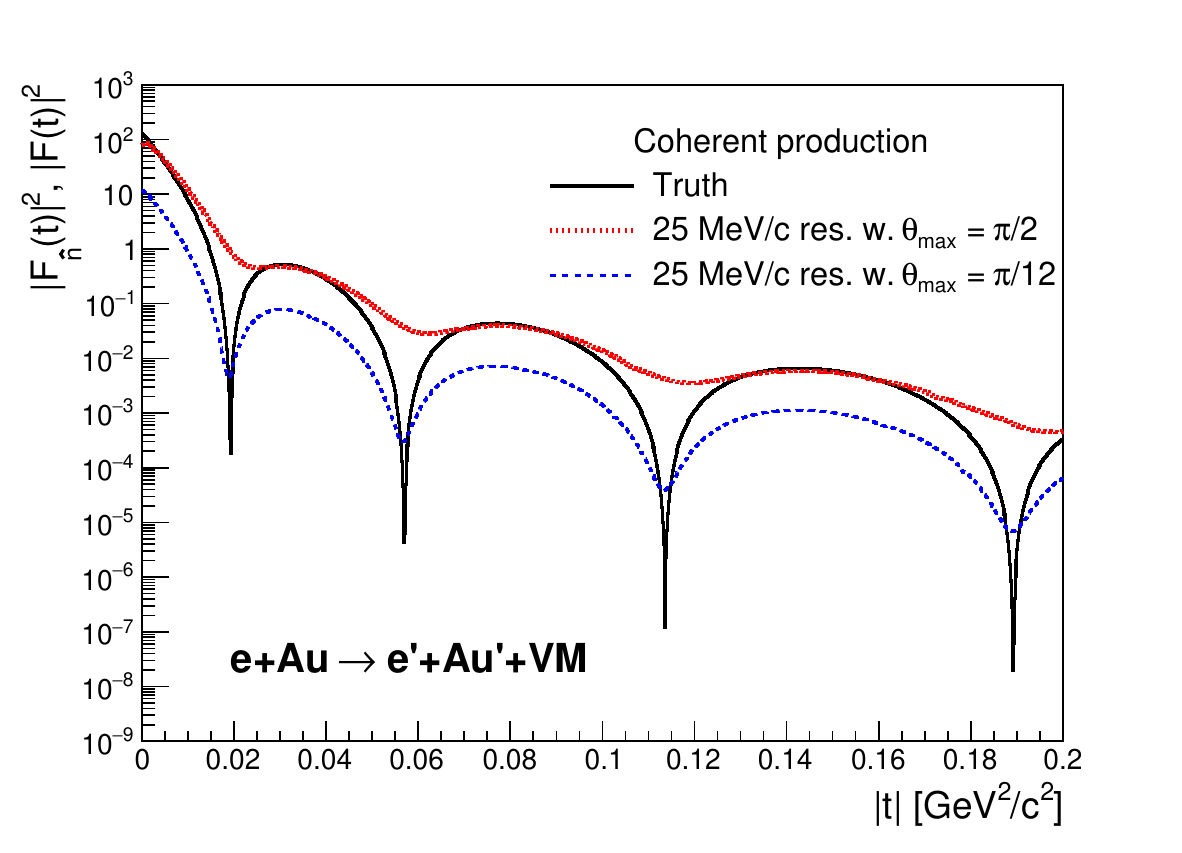}
    \caption
    {Projected $|t|$ distribution with a phase space constraint of $\theta_{\rm max} = \pi/12$ for coherent VM production. It is compared to the theoretical expectation of the full $|t|$ distribution as well as that with detector resolution but without the selection of phase space ({\it i.e.}, $\theta_{\rm max} = \pi/2$).
    }
    \label{fig:wedge_cut}
\end{figure}

The projected $|t|$ distribution ($|F_{\hat{n}} (t)|^2$) is obtained by applying a selection of $\theta_{\rm max} = \pi/12$, indicated by the black arrow in Fig. \ref{fig:ffq_25MeV}, and shown in Fig. \ref{fig:wedge_cut} as the blue dashed curve. Compared to the case without wedge cut or equivalently $\theta_{\rm max} = \pi/2$ (red dotted curve in Fig. \ref{fig:wedge_cut}), where the diffractive pattern is largely washed out, one can see a clear structure of peaks and valleys in the projected $|t|$ distribution for a wedge selection of $\theta_{\rm max} = \pi/12$. 
This significant enhancement in resolving the diffractive pattern is essential for using the $|t|$ distribution to image 
the gluon spatial distribution inside nuclei. However, it is worth noting that a loss of statistics occurs when employing a selective phase space as seen in Fig. \ref{fig:wedge_cut}. For example, about 83\% of coherent VM events are excluded with the $\theta_{\rm max} = \pi/12$ wedge cut for the 25 MeV/$c$ resolution. The optimal wedge selection should be determined experimentally based on the detector resolution and available statistics. For completeness, the analytical distribution of the form factor is also in Fig. \ref{fig:wedge_cut} as the black solid curve. As expected, the diffractive patten is more distinct compared to the case with detector resolution and without a phase-space selection in the specific $|t|$ decomposition. 

%
%

\section{Summary and Outlook}
\label{sec:}
In summary, we developed a unique approach to overcome the challenges of overwhelming incoherent background and limited detector resolution in utilizing coherent VM production for imaging the gluon spatial distribution inside nuclei. Specifically, the incoherent background can be separated from coherent VM production in collisions of transversely polarized electron beams by exploiting the angular distribution of decay daughters from VM. Moreover, the projected $|t|$ distribution perpendicular to the electron scattering plane can be used to avoid the effect of limited precision in measuring the outgoing electron's momentum. This is achieved by applying a selective phase space constraint in the transverse $|t|$ plane. Consequently, the theoretically predicted diffractive pattern for coherent VM production is largely restored, enabling a high-priority program at the EIC and other facilities aimed at measuring the gluon distribution in the nucleus and its dynamics.

In future work, we will implement this approach into the simulation of the ePIC detector, the baseline detector for the EIC. We will determine the phase-space selection based on the projected EIC luminosity and realistic angular resolution of outgoing electrons. An unfolding procedure will be employed to correct for the detector resolution in the measured projected $|t|$ distribution. The proposed technique can also be applied to other electron-hadron experiments, such as those in the Continuous Electron Beam Accelerator Facility at JLab~\cite{paolone_2014,paolone_2016}, EicC in China~\cite{Anderle:2021wcy} and LHeC at CERN~\cite{agostini2021large,LHeC2022}, to fully exploit the power of imaging gluon distributions with coherent VM production. 
This research is supported by the US Department of Energy, Office of Nuclear Physics (DOE NP), under contract Nos. DE-FG02-89ER40531, DE-SC0012704.

\biboptions{numbers,sort&compress}
\bibliographystyle{elsarticle-num} 
\bibliography{EIC_VM}
\end{document}